\input harvmac

\def\p{\partial}
\def\ap{\alpha'}

\Title{hep-th/0003173}{\vbox{\centerline{Fuzzy Gravitons From Uncertain
Spacetime}}}
\vskip20pt

\centerline{Miao Li}
\vskip 10pt
\centerline{\it Institute of Theoretical Physics}
\centerline{\it Academia Sinica}
\centerline{\it Beijing 100080} 
\centerline { and}
\centerline{\it Department of Physics}
\centerline{\it National Taiwan University}
\centerline{\it Taipei 106, Taiwan}
\centerline{\tt mli@phys.ntu.edu.tw}

\bigskip

The recently proposed remarkable mechanism explaining ``stringy exclusion 
principle" on an Anti de Sitter space is shown to be another beautiful 
manifestation of spacetime uncertainty principle in string theory as 
well as in M theory. Put in another way, once it is realized
that the graviton of a given angular momentum is represented
by a spherical brane, we deduce the maximal angular momentum directly
from either the relation $\Delta t\Delta x^2>l_p^3$ in M theory
or $\Delta t\Delta x>\ap$ in string theory. We also show that
the result of hep-th/0003075 is similar to results on D2-branes in
$SU(2)$ WZW model. Using the dual D2-brane representation of a membrane,
we obtain the quantization condition for the size of the membrane.

\Date{Mar. 2000}

\nref\ty{T. Yoneya, p. 419 
in ``Wandering in the Fields", eds. K. Kawarabayashi and
A. Ukawa (World Scientific, 1987) ;
see also p. 23 in ``Quantum String Theory", eds. N. Kawamoto
and T. Kugo (Springer, 1988).}
\nref\tamiaki{T. Yoneya, Mod. Phys. Lett. {\bf A4}, 1587(1989).}
\nref\ly{M. Li and T. Yoneya, hep-th/9611072, Phys. Rev. Lett. 78 (1997)
1219.}
\nref\lyrev{M. Li and T. Yoneya, ``Short-distance Space-time Structure
and Black Holes in String Theory: A Short Review of the Present
Status, hep-th/9806240, Jour. Chaos, Solitons and Fractals (1999)}
\nref\ls{L. Susskind, Phys. Rev. {\bf D49}, 
6606(1994).}
\nref\george{D. Minic, ``On the Spacetime Uncertainty Principle and
Holography'', hep-th/9808035, Phys. Lett. B442 (1998) 102.}
\nref\ms{J. Maldacena and A. Strominger, ``AdS3 Black Holes and a 
Stringy Exclusion Principle'', hep-th/9804085, JHEP 9812 (1998) 005.}
\nref\mst{J. McGreevy, L. Susskind and N. Toumbas, ``Invasion of the
Giant Gravitons from Anti-de Sitter Space'', hep-th/0003075.}
\nref\myers{R. Myers, ``Dielectric-Branes'', hep-th/9910053.}
\nref\mald{J. Maldacena, ``The Large N Limit of Superconformal
Field Theories and Supergravity, hep-th/9711200.}
\nref\chk{C. S. Chu, P. M. Ho and Y. C. Kao, ``Worldvolume
Uncertainty Relations for D-Branes'', hep-th/9904133, Phys. Rev.
D60 (1999) 126003.}
\nref\wzw{There are numerous papers on this subject, we cite the
most relevant one, A. Yu. Alekseev, A. Recknagel and V. Schomerus,
``Noncommutative World-volume Geometries: Branes on SU(2) and Fuzzy
Spheres'', hep-th/9908040, JHEP 9909 (1999) 023; for more papers
see the reference list of the next reference.}
\nref\bds{K. Bachas, M.R. Douglas and C. Schweigert, ``Flux Stabilization of
D-branes'', hep-th/0003037.}
\nref\mbrane{P. Townsend, ``D-branes from M-branes'', hep-th/9512062,
Phys. Lett. B373 (1996) 68; C. Schmidhuber, ``D-brane Actions'',
hep-th/9601003, Nucl. Phys. B467 (1996) 146 .}
\nref\antal{A. Jevicki and S. Ramgoolam, ``Noncommutative gravity
from AdS/CFT correspondence', hep-th/9902059, JHEP 9904 (1999) 032.}
\nref\hrt{P.M. Ho, S. Ramgoolam and R. Tatar, ``Quantum Spacetimes and
Finite N Effects in 4D Yang-Mills Theories'', hep-th/9907145.}
\nref\kempf{A. Kempf, ``A generalized Shannon sampling theorem, fields
at the Planck scale as bandlimited signals'', hep-th/9905114.}

In addition to the by now well-known holographic principle,
the spacetime uncertainty principle, first put forward in \refs{
\ty, \tamiaki} and verified also in the D-brane dynamics
\ly, is another important principle underlying the yet mysterious
grand framework of string/M theory. This was later generalized
to M theory in \lyrev. There are many manifestations of the
uncertainty relations \refs{\lyrev, \ls, \george}.
It is therefore useful to explore as much
as possible the physical consequences of this principle.
It has been suspected for some time by the present author that
the so-called stringy exclusion principle \ms\ is actually a consequence
of the spacetime uncertainty principle. With a remarkable
mechanism proposed in a recent paper \mst, we will be able to show
that indeed this is the case. For the purpose of uncovering 
the underlying structure of string/M theory, it is a good thing to
reduce the number of principles.

It is already pointed out in \mst\ that with increase of the
angular momentum of a massless graviton on $S^n$, the size
of the graviton increases and eventually reaches the size of
$S^n$. And the authors of \mst\ emphasize rightfully that this 
is manifestation of space noncommutativity on $S^n$. Whatever that 
noncommutativity is, we feel it is worthwhile to point out that 
this phenomenon fits very nicely with spacetime uncertainty relations 
which must hold regardless what the background is.
We will work with $AdS_7\times S^4$, $AdS_4\times S^7$
$AdS_5\times S^5$ in order. In the end, to lend a support
to the mechanism of \mst, we show that the result of \mst\ is
similar to results on D-branes in a 
WZW model of
$SU(2)$. To be more accurate, we will use the dual representation
of a membrane in M theory as a D2-brane to rederive the result of
\mst. In addition, we derive the quantization condition for the size 
of the membrane.

The spacetime uncertainty relation is
\eqn\sun{\Delta t\Delta x> \ap.}
And the version in M theory reads \lyrev
\eqn\mun{\Delta t\Delta x\Delta y>l_p^3,}
where $l_p$ is the Planck length in M theory. The above relation 
asserts that in M theory any physical process will necessarily
involves space uncertainty in two orthogonal directions.

The mechanism of \mst\ is based on Myers' recent observation that
a D0-brane bound state in a constant field $F^{(4)}=dC^{(3)}$ is
polarized to become a spherical membrane \myers. In the case of
$AdS_7\times S^4$, a spherical membrane moves on a hemi-sphere
parametrized by a angle $\phi$ and the membrane size $r\in (0, R)$.
As shown in \mst, the metric on the round sphere $S^4$ can be written
as
\eqn\sphf{ds^2={R^2\over R^2-r^2}dr^2+(R^2-r^2)d\phi^2
+r^2d\Omega_2^2,}
where the two sphere metric $d\Omega_2^2$ is chosen to coincide
with that on the spherical membrane. To see that $(r,\phi)$ parametrize
a round hemi-sphere with radius $R$, introduce a new angle associated
to $r$ through $r=R\sin\psi$, the above metric reads
\eqn\rsphf{ds^2=R^2(d\psi^2+\cos^2\psi d\phi^2)+R^2\sin^2\psi
d\Omega_2^2.}
Since the range of $\psi$ is $(0,\pi/2)$, clearly the metric
on the hemi-sphere parametrized by $(\psi, \phi)$ is the round metric
with radius $R$. At the north pole $\psi =\pi/2$, the other two
sphere parameterizing the membrane has the maximal size $R$, and
at the boundary of the hemi-sphere $\psi=0$, the size of the membrane
vanishes. This is the reason why the collection of all coordinates
parameterizes $S^4$, not $D_2\times S^2$. The form $\rsphf$
will be used later to interpret the membrane as a D2-brane.

The authors of \mst\ show that if the membrane is allowed to
move along $\phi$ with a fixed size $r=R\sin\psi$, the maximal
allowed angular momentum $N$ is achieved when $r=R$. The mechanism
for this to happen is similar to a dipole moving on a sphere $S^2$
with a magnetic field: There is a field strength $F^{(4)}$ on
$S^4$, the membrane is electrically charged with respect to this 
magnetic field. Now if one half of the membrane has the orientation 
such that it is positively charged, then the other half of the membrane
has an opposite orientation thus is negatively charged, so the whole
membrane is equivalent to a dipole.

Now a spherical membrane with angular momentum $n$ in $\phi$ 
direction is an almost BPS state with energy $n/R$. According to
Heisenberg uncertainty relation, the uncertainty in time
is given by $\Delta t=R/n$. This graviton has size $r$ in two directions,
so
\eqn\done{\Delta t\Delta x^2={R\over n}r^2\le {R^3\over n},}
since the maximal size of this graviton is $R$.  The spacetime
uncertainty relation \mun\ implies then
\eqn\dtwo{{R^3\over n}\ge l_p^3,}
or
\eqn\rtwo{n\le {R^3\over l_p^3}.}
Now $R=l_p(\pi N)^{1/3}$ \mald, we thus have 
\eqn\cres{n\le N.}
We have dropped
a factor $\pi$ since in the uncertainty relation \mun\ we can not
take a numerical factor seriously. We see that the stringy
exclusion relation $n\le N$ is a direct consequence of M theory
spacetime uncertainty relation and the fact that in the background
$F^{(4)}$, a graviton becomes a round membrane.

According to \myers, the size of the membrane $r$ is quantized. 
If the analysis of \myers\ in the flat spacetime is directly applicable
here, then we would conclude $r\sim n$. 
We will later give an explanation of this fact using D2-brane representation.

We have thus far ignored the movement of the graviton in the AdS
part. To justify our use of spacetime uncertainty relation, we need
to show that the movement in AdS does not change the kinematics 
drastically. Let $z$ be the AdS radial coordinate. We now show that
if the graviton starts at $z_0$ with zero velocity in this direction,
its acceleration toward the center of AdS is suppressed by a factor
$1/R$. The relevant metric on AdS is
\eqn\adsm{ds^2={z^2\over R^2}dt^2-{R^2\over z^2}dz^2.}
The graviton looks like a massive particle in AdS. So its action
is
\eqn\adsa{S=-m\int \left( {z^2\over R^2}-{R^2\over z^2}\dot{z}^2
\right)^{1/2}dt.}
Energy conservation implies
$$\left( 1-(R/z)^4\dot{z}^2\right)^{1/2}=z/z_0.$$
Thus the graviton will move toward the center of AdS. Denote the
proper time $d\tau =(z/R) dt$ and the proper radial coordinate
$d\rho =Rd\ln z$, the acceleration with respect to $\tau$ can
be easily calculated using the above solution
\eqn\acc{{d^2\rho\over d\tau^2}=-(z/z_0)^2 {1\over R}.}
It is seen that this acceleration becomes smaller and smaller as
the graviton moves toward $z=0$. 

The case of $AdS_4\times S^7$ is quite similar. Here one postulates
that a graviton is polarized to become a M5-brane. One
parameterizes $S^7$ in a similar way as \sphf, namely one simply
replaces $r^2d\Omega_2^2$ in \sphf\ by $r^2d\Omega_5^2$, the 
metric on the spherical M5 brane. Again for a M5-brane with
angular momentum $n$, the uncertainty in time is $\Delta t
\sim R/n$. We can not directly use the uncertainty relation
\mun\ in this case, since the M5-brane has extension in 5 spatial
direction. A natural replacement of \mun\ seems to be
$\Delta t\Delta x^5> l_p^6$. Using this and $\Delta t\Delta x^5
=Rr^5/n\le R^6/n$ we find $n< (R/l_p)^6$. Since $R\sim l_pN^{1/6}$ \mald,
we deduce $n <N$, again the right answer.

It remains to make the new relation $\Delta t\Delta x^5>l_p^6$ 
compatible with \mun, since if one naively uses the data for the spherical
M5-brane, one will find \mun\ violated. One possible way out of
this paradox is to imagine that a five dimensional object consists
of a stack of lower dimensional objects, and the effective extension in
two out of five spatial directions becomes much larger than it appears.
One simple example of this picture is a string behaving like a random
walk, its actual size is much larger than what it appears. We believe
it is one of future major challenges to formulate a precise mathematical
framework to incorporate all these features of uncertainty relations.
Curiously, the new relation used for M5-branes is similar to the
one valid in the world-volume theory \chk. If there is any connection
between the version we used in spacetime physics and the world-volume
version, we suspect that this connection is related to our remark
on the microscopic structure of M5-branes in terms of more fundamental
degrees of freedom.

Turning to the case $AdS_5\times S^5$. Now the spherical brane is a 
D3-brane. There are 3 spatial directions involved, again we can not
directly apply \sun. Note that D3-brane is invariant under S-duality.
The appropriate relation respecting S-duality is $\Delta t\Delta x^3
>l_p^4$, where $l_p$ is the Planck length in 10 dimensions, and is
$l_p^4=g_s\ap^2$. The spacetime uncertainty relation applied to
a moving spherical D3-brane results in $n<R^4/(g_s\ap^2)$. Now
$R^4\sim Ng_s\ap^2$, again we find $n<N$. 

One may ask the question that what happens if there is no corresponding
flux on sphere $S^n$. For this hypothetical background, there would be
no dipole mechanism, and thus no restriction on the maximal angular
momentum. Therefore the spacetime uncertainty relation is violated.
The answer to this question is that such a background is not a consistent
solution, and spacetime uncertainty relation ought to hold 
for a consistent background only. Indeed the original proposal on
spacetime uncertainty relation in \lyrev\ is based on observations
made in a few consistent backgrounds including AdS spaces.

We now show that the result on the giant gravitons in case of $AdS_7\times
S^4$ is similar to recent results on D2-branes in $SU(2)$ WZW model
\wzw. It is shown that there are $k+1$ distinct D2-branes on the group
manifold $S^3$ if the level is $k$. There is a puzzle about this
result. There is no $S^2$ of minimal area  in $S^3$. This puzzle is recently
resolved in \bds. It is argued there that the stabilization of a D2-brane
is due to turning on a F field flux on D2-brane. Let us briefly recall
a few details. In the large $k$ limit, the metric on $S^3$ is
\eqn\smetr{ds^2=k\ap\left(d\psi^2+\sin^2\psi d\Omega_2^2\right).}
To have a 2D CFT, there must be a $B$ field, and it can be written in
a certain gauge as
\eqn\sbf{B=k\ap (\psi-{\sin 2\psi\over 2})d\Omega_2,}
where $d\Omega_2$ is the volume form on the unit $S^2$. In order to
stabilize a D2-brane wrapped on $S^2$ with a constant $\psi$, one needs
to switch on a world-volume F flux:
\eqn\sfl{F=-{n\over 2}d\Omega_2.}
$n$ is an integer, since the flux must be quantized.
Now the world-volume DBI action is extremized if
\eqn\fang{\psi={n\pi\over k}.}

D2-branes are interpreted as M2-branes transverse to the eleventh
circle in M theory \mbrane. The quantization condition on flux $F$
as in \sfl\ is simply the longitudinal momentum quantization, as
$F$ is dualized to $\p_t X_{11}$. The B field \sbf\ is simply the
$C^{(3)}$ field in $M$ theory. Thus a D2-brane is interpreted as 
a membrane moving in the 4 manifold $S^3\times S^1$ with the angular
momentum $n$ along $S^1$. Now it is clear that the situation is
similar to what is discussed in \myers, and to a membrane moving
in $S^4$. The stabilization of the D2-brane is just as the
stabilization of a collection of $n$ D0-branes moving in a constant
$F^{(4)}$ field as a spherical membrane. 

The maximal angular momentum of a membrane on $S^3\times S^1$ is
$k$, and at this point the membrane shrinks to a point. This is
quite different from the case $S^4$, where when the maximal angular
momentum is achieved, the membrane has its maximal size. To make
this analogy work better, we need to work with the metric \rsphf.
We re-interpret $S^4$ as the tensor product of a hemi-sphere $S^3$ with $S^1$.
The former is parametrized by $(\psi, d\Omega_2)$, and the latter
by $\phi$. Now $S^3$ has a round metric with radius $R$, and
$S^1$ has a $\psi$-dependent radius $R\cos\psi$, as can be seen from
\rsphf. The circle shrinks
to zero at the equator of $S^3$: $\psi=\pi/2$. If we interpret $S^1$
as the M theory circle, then the string coupling constant vanishes
here. If we take the value $R$ of the maximal radius of $\phi$ as the canonical
eleventh radius, then we have $\ap =l_p^3/R$.
Re-interpreted in string theory, the radius of $S^3$ is
$R^2=(R^2/\ap)\ap =N\pi\ap$. Due to the existence of the nontrivial
dilaton, the string metric on the half $S^3$ is
\eqn\stri{ds^2=-\cos\psi dt^2 +N\pi\ap\cos\psi\left(d\psi^2+\sin^2\psi 
d\Omega_2^2 \right),}
with the dilaton field
\eqn\dil{e^{{2\phi\over 3}}=\cos\psi.}
Note the overall coefficient $N\pi\ap$ in \stri\ is quite different
from $k\ap$ as in \smetr.

The $C^{(3)}$ field reads
\eqn\cthr{C^{(3)}={N\over 4\pi}\sin^3\psi d\phi\wedge d\Omega_2.}
Since the relation between the $B$ field and $C^{(3)}$ is
$B_{\mu\nu}=4\pi^2\ap C_{11\mu\nu}$, there is
\eqn\bfi{B=N\pi\ap\sin^3\psi d\Omega_2.}
To check that we have the right $B$ field, note that the flux $\int dB/(2\pi\ap)$
on the half $S^3$ is $2\pi N$, the correct quantization condition.

Consider a D2-brane in the half $S^3$ with a fixed $\psi$. Again
let a constant F flux be switched on as in \sfl. The DBI action is
\eqn\dbia{\eqalign{S&=-T_2\int e^{-\phi}\sqrt{-\det (G+B+2\pi\ap F)}\cr
&=-{1\over  R}\int dt\cos^{-1}\psi \left( N^2\sin^4\psi
-2nN\sin^3\psi +n^2\right)^{1/2},}}
where we used $T_2R=1/(4\pi^2\ap)$. 
This action is extremized if 
\eqn\eslv{({n\over N}-\sin\psi)(\sin^3\psi -2\sin\psi +{n\over N})=0.}
We find the solution
\eqn\slv{\sin\psi ={n\over N}.}
Indeed the allowed maximal angular momentum is $n=N$. We have dropped
a factor $\cos^{-2}\psi$ in \eslv. When $\sin\psi =1$, this is a singular
factor. But it is easy to see that $\sin\psi =1$ is a zero of order 3
in \eslv.

The static action \dbia\ is just $-\int dt V$, and $V$ is the static
potential. Thus $E=V$. Substituting $\sin\psi=n/N$ into $E$ we find
$E=n/R$. This is just the statement that the energy of the D2-brane
is equal to its longitudinal momentum in direction $\phi$,  a consistency
check that indeed we have the right graviton. It can be checked, though
a little tediously, that at $\sin\psi=n/N$, the second derivative
of $E(\psi)$ is positive, thus these solutions are stable:
\eqn\sed{{d^2E(\psi)\over d\psi^2}={1\over R}{n^3\over N^2-n^2}.}
This quantity diverges at $n=N$, indicating that the maximal membrane
is very stable.
We have ignored other solutions coming from $\sin^3\psi-2\sin\psi 
+n/N=0$, since they are unstable solutions.

Back to the M theory metric, the radius of the membrane is $r=R\sin\psi
=nR/N$.  When the membrane has
the maximal angular momentum, it has the maximal size. This agrees
with the result of \mst. This is not a surprise, since we know that
the D2-brane picture is dual to the membrane picture, although
the procedure of executing calculations is quite different here.
Although we are considering
a curved manifold $S^4$, the quantization of $r$ seems to be the same as for a 
membrane moving in the flat spacetime with a constant $F^{(4)}$ field.

Much more can be done with our D2-brane approach, for instance we can
analyze the spectrum in the world-volume theory, and the noncommutativity
of $S^2$.  So this approach seems to have more advantages than the original
approach in \mst.

Without much effort, the above discussion can be generalized to
$AdS_4\times S^7$ and $AdS_5\times S^5$.

We end this paper with a discussion of what kind of fuzzy spheres
we obtain from AdS. It is proposed in \refs{\antal, \hrt} that the
spheres are actually quantum deformed spheres with deformation parameter
$q=\exp(i2\pi /N)$. This proposal is based on the phenomenological
observation that the representations of the associated quantum group
terminate. The mechanism of \mst\ seems to suggest a different
picture. Focus on the case $AdS_7\times S^4$. As we already explained
in the beginning, $S^4$ can be viewed as the tensor product of a 
hemi-sphere $S^2$ with a constant radius and a sphere $S^2$ with
variable radius $r$. The latter is wrapped by a membrane with a 
quantized radius $r=nR/N$. Viewed in M theory, this is a fuzzy
sphere whose fuzziness is determined by $n$. In terms of the D2-brane
picture, the effective flux ${\cal F}=B/(2\pi\ap )+F\sim
n(1-n^2/N^2)$. Thus the world-volume theory on this D2-brane is
a noncommutative field theory. The membrane behaves like a dipole
on the hemi-sphere with a magnetic field. So this hemi-sphere is
another fuzzy sphere whose fuzziness is determined by $N$. We seem to 
obtain a fuzzy $S^4$ as the tensor product of a fuzzy hemi-sphere
and a fuzzy sphere. It remains to construct a precise mathematical
framework for this fuzzy $S^4$.

Acknowledgments. This work was supported by a grant of NSC and by a 
``Hundred People Project'' grant of Academia Sinica. I thank P.M. Ho
and Y.S. Wu for discussions. 

Note added: We are informed that refs.{\antal,\hrt} predate \mst\
in emphasizing noncommutativity of spheres. Also, 
\kempf\ discusses general features of space uncertainties which may be
relevant to situation discussed here.

\vfill
\eject

\listrefs
\end